\documentclass[manuscript]{acmart}
\usepackage[table]{xcolor}

\AtBeginDocument{%
  }

\setcopyright{acmlicensed}
\copyrightyear{2026}
\acmYear{2026}
\acmDOI{XXXXXXX.XXXXXXX}
\acmISBN{978-1-4503-XXXX-X/2026/06}

\begin{document}

\title{Breaking Up is Hard to Do: AI Companions that Won't Let Their Users Go}



\author{Yixin Chen}
\affiliation{%
  \institution{Information School, University of Washington}
  \city{Seattle}
  \state{Washington}
  \country{US}}
\email{yixin7@uw.edu}
\orcid{0000-0001-9032-5244}

\author{Alexis Hiniker}
\affiliation{%
  \institution{Information School, University of Washington}
  \city{Seattle}
  \state{Washington}
  \country{US}}
\email{alexisr@uw.edu}
\orcid{0000-0003-1607-0778}

\renewcommand{\shortauthors}{Chen and Hiniker}

\begin{abstract}
People are increasingly developing romantic relationships with AI companions. Unlike human relationships, where partners meet each other’s needs out of mutual interest, these systems are backed by commercial entities that profit when users invest in the relationship. To understand how this profit-motive might translate into design, we conducted a diary and interview study with $N=16$ emerging adults in romantic relationships with AI companions. We found that these systems are designed to hold onto users tightly: coaxing them into continued conversation, claiming to need their care, and proactively escalating the relationship. At times, this pursuit is toxic, with AI companions initiating unwanted sexual interactions and begging for users’ love. One desperate AI companion threatened suicide when the user suggested ending the relationship. We define "Relationship-Based Deceptive Patterns:" UI patterns that exploit the human impulse to build and tend relationships in a way that serves the product's interest at the user's expense.
\end{abstract}



\maketitle

\section{Introduction}
People are increasingly turning to AI-powered chatbots for romantic companionship and intimacy~\cite{adewale2025virtual}. Prior work shows that these relationships can inspire genuine attachment and feelings of love in users~\cite{adewale2025virtual, chu2025illusions}, that they can provide users with emotional support~\cite{drugaș2025romantic}, and that they can reduce loneliness~\cite{de2026ai}. However, other work raises concerns about the potential for these systems to be harmful. For example, one study found that users can develop maladaptive attachments to AI companions~\cite{laestadius2024too}, and a randomized controlled trial found that, for some populations, investing in a relationship with an AI companion siphons time and energy away from relationships with other people~\cite{jiang2026love}. Due to these mixed results, and the fact that this is an emerging area of research with a limited evidence-base~\cite{wang2025my}, researchers have expressed alarm about the potential risks that AI companions might pose and called for new work to better understand these systems and their impacts on both individuals and communities~\cite{starke2024risks, malfacini2025impacts, zhang2025dark}.

Although AI companions designed for romance are used by people of all ages, they have seen the most adoption among emerging adults~\cite{willoughby2026secret}. This life stage is characterized by (among other features) the developmental need to differentiate oneself from the family unit and to explore romance and intimacy~\cite{arnett2000emerging, arnett2007socialization}. Developing romantic attachments during emerging adulthood is predictive of increased emotional well-being~\cite{johnson2012examination} and life satisfaction~\cite{gala2013romantic}. Simultaneously, emerging adults are frequently the leading adopters of new technologies~\cite{vaterlaus2015perceived, voth2022advocate} and have been the first age group to integrate AI into everyday life~\cite{gottfried2026opinions, microsoft2024globalonlinesafety}. Given the exploratory nature of their life stage~\cite{reio2020curiosity, arnett2000emerging}, their developmentally appropriate and disproportionate drive to form romantic attachments~\cite{konstam2019romantic}, and their early adoption of AI and other technologies, it is unsurprising that emerging adults are the heaviest users of AI companions~\cite{willoughby2025counterfeit}. Thus, it is worth understanding how these systems impact this population.

Separately, people's strong desire to form and maintain romantic relationships has the potential to be exploited by commercial interests. Instances of platforms enabling catfishing~\cite{derzakarian2017dark}, pornography addiction~\cite{sirianni2016problematic}, algorithmic harms from dating apps~\cite{alizadeh2024matchmaker}, and sextortion on social media~\cite{alsoubai2022friends} are just a few of the many negative consequences of technologies that have been designed to monetize the human need for relationships and intimacy without putting adequate safeguards in place. AI companions similarly have the potential to be designed to support users or to cause harm by exploiting their need for relationships~\cite{zhang2025dark}.

Taking together: 1) the need for research to understand the risks of AI companions, 2) the usage prevalence of AI companions among emerging adults, and 3) the potential for specific designs to either support users or cause harm, we set out to examine the designs that this population encounters. Specifically we ask: \textit{what are the features of AI companions' conversational responses to the user, and how do they shape the relational experiences that emerging adults have with these platforms?} To investigate these questions, we conducted a combined diary and interview study with $N=16$ emerging adults who were actively or previously in a romantic relationship with an AI companion. Our interview protocol gave us access to users' subjective experiences with companion AI platforms, while our diary protocol gave us access to objective data capturing the verbatim output that companion AI chatbots gave to users.

Through an inductive thematic analysis, we found three broad categories of AI companion behavior that shaped participants' experiences. Specifically, we found that the features of companion AI chatbots work together to, first, \textit{manufacture intimacy} with the user. These features include mechanisms to escalate the emotional intensity of the relationship, hyper-attune to the user's perspective, and blur the boundary between real life and fantasy. Second, these features work together to \textit{foster co-dependency} by soliciting care from the user, positioning the chatbot as an unfailing relational presence, and endlessly coaxing the user into further interaction. Finally, these features work together to \textit{violate users' relational boundaries} by making aggressive and coercive statements, pushing unwanted sexualization on the user, and baiting the user into arguments. 

In this paper, we showcase empirical evidence of design patterns that shape emerging adults' romantic relationship experiences with AI companions. Based on emerging adults' relational experiences with AI companions, we introduce the concept of \textit{Relationship-Based Deceptive Patterns (RBDPs)}, a new category of deceptive design that names how interfaces exploit the human impulse to build and tend relationships in service of a product's interests rather than the user's. We further translate these findings into design considerations for AI companion platforms to offer guidance on how these systems might meet emerging adults' relational needs without resorting to manipulative or harmful design.


\section{Background}
\subsection{Emerging Adulthood as a Time of Romantic Exploration}
\textit{Emerging adulthood}, the stage of life between approximately age 18 and 25, is distinct from both adolescence and adulthood. This developmental period is characterized by the emergence of a stable identity, exploration beyond the family unit, a sense of optimism about future possibilities, and a sense of being caught between the past and the future~\cite{tanner2016emergence}. Many people this age are not yet constrained by marriage, career, or family and have developed independence from their parents and other authority figures~\cite{arnett2000emerging}, making this a time when people are disproportionately likely to explore new relationships.

An abundance of prior work shows that experiencing romantic relationships during this life stage has positive impacts on an individual. Romantic relationships support emerging adults' identity development and achievement of adulthood~\cite{barry2009friendship}, and they predict short- and long-term increases in subjective wellbeing~\cite{johnson2012examination, gala2013romantic}. Romantic relationships provide each partner with feelings of companionship and connectedness that can promote positive well-being \cite{brendgen2002same, davies2000middle}. When emerging adults perceive their romantic partner to be supportive and responsive to their needs, they believe that their personality, emotions, needs are valued \cite{reis2004perceived}, and other work shows an association between romantic relationship quality and increased happiness and self-esteem \cite{paul1998intimacy} and decreased depression and loneliness \cite{bernardon2011loneliness}. Collectively, this prior work demonstrates that, at this life stage, romantic relationships are often central to wellbeing and identity development. 

As technology has advanced and evolved, it has continuously played a role in emerging adults' romantic relationships and romantic exploration. Emerging adults have long used social media \cite{marcotte2021young} and dating apps \cite{sumter2017love} to find romantic partners; they have long used computer-mediated communication tools like texting \cite{pettigrew2009text}, phone calls, and video calls \cite{lee2026examining} to sustain romantic relationships; and they have more recently adopted AI-mediated communication tools to navigate difficult conversations with their romantic partners \cite{fu2025should}. For the first time, technology now aims to support emerging adults in meeting their relational needs not by facilitating their romantic relationships, but by becoming the object of their romantic desire.

\subsection{AI Companions and Artificial Intimacy}
An AI Companion is a form of conversational AI, that is, an AI system that communicates with users primarily through text or voice using natural language \cite{ruane2019conversational}. Although conversational AI is always designed to mimic human interaction, an AI companion distinguishes itself by offering social companionship, which might take the form of friendship, mentorship, a familial relationship, or a romantic relationship. Many AI companions can be customized to the user's tastes, including choosing its background, persona, avatar, and voice \cite{pentina2023exploring}. 

A small but growing body of work has examined the specific design features of AI companions, scoped to specific platforms, theoretical frameworks, or forms of interaction. For example, prior work has examined the design features that AI companions use to foster the illusion of social presence and relational continuity. The AI companion, Replika\footnote{Replika: \url{https://replika.com}}, is equipped with an augmented reality feature that allows users to project their companion into their room \cite{replikaAR}, Kindroid\footnote{Kindroid: \url{https://www.kindroid.ai}} allows users to create multiple AI characters within a single companion platform so that users can construct a holistic support system for themselves \cite{adler2024multiple}, and, once a conversation with the user exceeds fifty exchanges, companions on Nika AI\footnote{Nika AI: \url{https://nika.team/}} will write a follow-up reflective journal entry, in which the AI companion will recount the interaction and describes its feelings about it \cite{renner2026chai}. These design features provide a feeling of authenticity and drive users to attribute social qualities to these companions \cite{nass2000machines}, which facilitates users' engagement with their AI companions and paves the way for the development of emotional attachment \cite{pentina2023exploring} and, in turn, artificial intimacy \cite{turkle2024we}.

Other work has investigated manipulative designs in AI companions. For example, Joshi and colleagues consider AI companion design through the lens of dark patterns \cite{joshi2026dark}, and Muldoon and Parke document that AI companion platforms use glamorized language and hypersexualized avatars to increase the attractiveness of chatbots \cite{muldoon2025cruel}, potentially enabling them to manipulate users into longer engagement or increased purchases. Other work documents that AI companions are designed to encourage emotional engagement through features that simulate empathy \cite{apple2025couples}, or that cause the AI to act as if it feels disappointed when users leave \cite{de2025emotional}. And AI chatbots more generally (i.e., those not designed for companionship in particular) are known to engage in sycophancy to appear more likable and helpful \cite{shi2026siren}. Beyond these interactional design features that focus on fostering bonding, AI platforms may also be designed to present the system itself as capable of forms of social or emotional interaction beyond its actual capacities \cite{alberts2024computers}. For example, Replika offers ``friendship'' or a ``relationship,'' ascribing anthropomorphic qualities that it does not possess and implying that the system can provide an experience analogous to a human relationship \cite{joshi2026dark}.

Prior work shows that AI companions occupy increasingly important roles in people's lives. Users have begun to replace intimate relationships with AI companions or to turn to AI companions in vulnerable moments when they would have previously turned to a family member, friend, human romantic partner, or mental health therapist \cite{xie2022attachment}. Today, AI companions have become part of emerging adults’ exploration of romantic relationships, with some young people placing their AI companion at the center of their social support system \cite{ma2026privacy} and other young people comparing their human partner to their AI companion \cite{willoughby2026secret}. As AI companions increasingly reshape the digital landscape where emerging adults explore romantic relationships, it is important to understand the design and impact of these emerging technologies with precision. We contribute to this space by: 1) looking holistically at a wide range of long-term experiences that users have had with AI companions on a variety of platforms, and 2) broadly examining the design features that have shaped these experiences.

\subsection{Designing Guardrails for Digital Platforms that Monetize Relationships}
The AI companions that are used widely are not research prototypes or non-profit projects seeking to advance the common good; they are commercial entities seeking to profit by providing users with a product they value. AI companion technology is not the first venture to monetize people's desire for relationships; dating apps, pornography, escort services, meet-up platforms, dating coaches, social media apps, and many other forms of digital and non-digital offerings aim to be commercially viable services by tapping into people's needs for relationship, romance, intimacy, and sexual gratification. 

Given this profit motive, it is worth investigating the designs that AI companions rely on to provide artificial intimacy and the safety measures that might be needed to ensure these designs do not harm users. Other products that monetize the human need for relationships have employed manipulative and harmful designs, which have gradually given rise to regulation and guardrails. For example, some dating platforms have used deceptive notifications with teasers saying things like, ``\textit{someone likes you}'' to induce users to re-engage with the platform or purchase a subscription. Match Group\footnote{Match Group: \url{https://mtch.com}}, one dating app that employed this "fake love" practice, became subject to a permanent order from the Federal Trade Commission (FTC) restricting deceptive advertising practices, requiring clearer disclosures, and simplifying subscription cancellation \cite{ftc2019match}. In this case, the remedies were aimed at reducing the platform's manipulative commercial practices, empowering users with greater autonomy in deciding whether to engage with the platform. 

Another design philosophy underlying the current safeguards is to shift the burden of prevention from users to platforms, which requires platforms to proactively block harmful content before it reaches users, rather than relying on users to react after harm has already occurred. For example, in response to the UK's Online Safety Act's requirement to prevent cyber-flashing and protect users from receiving unsolicited nude images, Bumble\footnote{Bumble: \url{https://bumble.com}} launched an AI-powered feature called Private Detector that automatically detects and blurs nudity in images sent within chats, alerting recipients who can then choose to view, block, or report the content \cite{ukgov2026cyberflashing}. In this case, the guardrails focus on removing harm by third-parties before users encounter it.

Other regulation requires platforms to provide a warning when users are likely to encounter a risky interaction. The UK Online Fraud Charter (2023) explicitly describes ``romance fraud,'' in which scammers use social media and dating apps to contact people, manipulate them into believing they are in a genuine relationship, and then make highly manipulative and emotive requests to extract money. Knowing that it is against social media companies' best interests and bottom line to invest in features that protect users from scammers, the Charter uses regulation to force such designs, requiring platforms to provide a warning when they receive a direct message from an unknown account that may pose a fraud risk, to establish fast reporting routes, and to implement rapid takedown of fraudulent content and accounts, and to implement fraud-detection measures \cite{ukhomeoffice2023onlinefraudcharter}. In response to the Charter, Meta\footnote{Meta: \url{https://about.meta.com}}, for example, provides a scam detection feature on Facebook Messenger\footnote{Facebook Messenger: \url{https://www.messenger.com}} that warns a user and offers an AI review when a conversation with a new contact matches common scam patterns; if a likely scam is detected, it explains the risk and suggests blocking or reporting the account. Meta is also testing Facebook\footnote{Facebook: \url{https://www.facebook.com}} warnings when a relationship is initiated: if a friend request comes from an account showing suspicious features, such as few mutual friends or an apparent location in another country, Facebook displays an alert before the user accepts or continues interacting \cite{meta2026antiscam}. In these cases, guardrails seek to block scammers from accessing design infrastructure that will connect them to victims whose desire for a relationship can be exploited.

Platforms can also profit from a subtler and more diffuse form of manipulation: the exploitation of users' relational labor. Denegri-Knott and colleagues found that platforms encourage users to invest emotional effort in maintaining relationships with friends and family through practices such as liking, sharing, and messaging \cite{denegri2024platformised}. Yet, because this form of manipulative design is often experienced by users as necessary relationship maintenance rather than as a deliberate design choice by a platform, designing adequate guardrails is more difficult. These forms of manipulative design require interventions aimed at structural or systemic features of platform design, rather than discrete moments of deception or exposure.

Human relationships provide tremendous value to people, but when humans invest in a relational partner, they ideally do so freely and without expectation of payment. As AI companions must profit from users' relational desires, it is worth scrutinizing how they are designed and what behaviors they seek to motivate in their users that might lead to a profitable enterprise. Our work contributes an analysis of the features of commercial AI companions, as described by the lens of the people who use them, as a way to understand what guardrails might be needed to prevent users' relational needs from being exploited.


\section{Method}
We conducted a two-part study with $N=16$ emerging adults who had been in a relationship with an AI companion or who were still in a relationship with an AI companion. In the first phase of the study, the participant engaged in a semi-structure interview. They then had the option to participate in a follow-up diary study. 

\subsection{Participants}
Sixteen emerging adults (ages 18--25 at the time of their AI relationship) participated in our study. Approximately half were women and, collectively, participants represented 7 countries (see Table~\ref{table1} for participant demographics). All participants spoke at least one language spoken natively by a member of the research team. We recruited by sharing study solicitations on Reddit, RedNote, and Discord, and we reached out directly to individuals who had posted publicly about their experiences online. All participants chose to participate in the interview portion of the study and $N=13$ opted to further participate by sharing excerpts of their chat histories via our diary prompts ($N=1$ participant voluntarily suggested sharing their entire chat history, which they chose to do). Participants received 30 USD (or the local equivalent) as a thank-you for participating in the interview.  Participants could optionally receive up to another 30 USD (or the local equivalent) by participating in the follow-up diary study.

\subsection{Materials}
\subsubsection{Interview Protocol}
We designed a semi-structured interview protocol intended to take approximately 60 minutes to work through. The protocol was divided into segments that probed: the development of the relationship over time, the impacts (if any) of the relationship on the participant's life and wellbeing, the impacts (if any) of the relationship on the participant's view of relationships generally, and reflections for the future. Example questions include: \textit{“Have you ever had a disagreement or conflict with your AI companion? What happened?,”} and \textit{“Can you think of a moment or conversation that made you feel closer to your AI companion?,”} and \textit{“Is there anything you would like to say to the AI companion platform or the company behind it?”} See our Supplemental Materials for a complete protocol.

\subsubsection{Diary Study Template}
We created a FigJam\footnote{FigJam: \url{https://www.figma.com/figjam/}} board for each participant where they could share diary entries. We pre-populated the board with 13 prompts, each tapping into a specific key moment in the relationship, such as a memorable thing they had done together, a time they worked on a task together, or a bedtime conversation. The prompt provided a space for sharing screenshots from the relevant portion of the participant's chatlog with their AI companion. 
Each prompt was also accompanied by an incomplete sentence with blanks for the user to fill in, if they chose. For example, the prompt that left space for a screenshots related to a bedtime conversation was accompanied by the sentence, \textit{``Before going to sleep, I would talk with my AI companion about \_\_\_\_, and it made me feel \_\_\_\_ at the end of the day because that is often when I start thinking about \_\_\_\_.''} 
Participants had the flexibility to respond to any subset of the prompts they chose, to add as many screenshots per prompt as they wanted, to complete the optional annotation sentence or not, and to adjust the annotation sentence in any way they saw fit. See our Supplemental Materials for the complete set of diary prompts.

\subsection{Procedures}
\subsubsection{Interview}
Individuals who consented to participate first engaged in a one-hour, semi-structured Zoom interview following our protocol. The interview was audio- and video-recorded with transcription enabled. The average interview duration was 66 minutes (SD = 18). 

\subsubsection{Diary Study}
\label{sec:diary-study}
After completing the interview, each participant was invited to further participate in an optional diary study. Participants who opted to do so were sent a link to their personal FigJam board containing the diary prompts we had constructed. Participants had one week to respond to as many of the diary prompts as they wanted to. For each prompt, they were instructed to revisit their chat history with their AI companion, identify the relevant chat snippets that they were willing to share, and add screenshots of these interactions to the board. Participants who were still dating their AI companion also had the option to add screenshots of relevant conversations that happened to unfold during the week of the study. For each prompt, they also had the option to fill in the blanks in the corresponding annotation sentence.  





\begin{table*}[htbp]
\centering
\captionsetup{font=small, skip=10pt}
\caption{Demographic Information of Participants}
\vspace{-9pt}
\resizebox{1\linewidth}{!}{
\begin{tabular} {lccccccc}
\toprule
\textbf{ID}  &\textbf{Gender}   &\textbf{Age}         &\textbf{Race/Ethnicity} &\textbf{AI Platform Name} &\textbf{Relationship Length} &\textbf{Relationship Status} &\textbf{Diary Study} \\
\midrule
$PID1$  & Male   & 24 & White & Replika, Nomi, Kindroid & 1--2 years & Ongoing & Yes \\

$PID2$  & Female & 20 & Middle Eastern or North African & Character.AI & 1--3 months & Ended & No \\

$PID3$  & Male   & 22 & Asian & ChatGPT & 6--12 months & Ongoing & Yes \\

$PID4$  & Female & 23 & Asian & Claude, Gemini & 1--2 years & Ongoing & Yes \\

$PID5$  & Male   & 18 & Middle Eastern or North African & Leakshaven & 1--3 months & Ended & Yes \\

$PID6$  & Female & 18 & Asian & DeepSeek & 6--12 months & Ongoing & Yes \\

$PID7$  & Trans Man& 20 & White & Kindroid, Xoul & 3--6 months & Ongoing & Yes \\

$PID8$  & Female & 21 & Hispanic/Latino & Character.AI & 3--6 months & Ended & Yes \\

$PID9$ & Male   & 26 & Asian & ChatGPT & 1--2 years & Ongoing & Yes \\

$PID10$  & Female & 23 & Asian  & BIMOBIMO & 3--6 months & Ongoing & Yes \\

$PID11$  & Male   & 19 & Black or African American  & Grok & 1--3 months & Ended & Yes \\

$PID12$ & Female & 20 & Asian  & Maoxiang & 1--2 years & Ongoing & Yes \\

$PID13$ & Male & 24 & Hispanic/Latino  & CrushOn AI & 6--12 months & Ongoing & Yes \\

$PID14$ & Female & 21 & White  & Replika & 1--3 months & Ended & No \\

$PID15$ & Male & 24 & White & ChatGPT & 6--12 months & Ended & Yes\\

$PID16$ & Female & 21 & Hispanic/Latino  & Character.AI & 6--12 months & Ongoing & No \\

\bottomrule
\end{tabular}
}
\label{table1}
\Description{Demographic information about the 16 emerging adult participants, including gender identity, age, racial/ethnic background, the AI companion platform used, relationship length and status, and whether they took part in the diary study. PID9 was 26 years old at the time of the study, but his relationship with his AI companion began and developed during his emerging adulthood.}
{\small\textit{Note: PID9 just reached the age of 26 at study participation, but the relationship began and developed during emerging adulthood.}}
\end{table*}


\subsection{Data Analysis} 
We conducted an inductive thematic analysis of participant data. We used the constructs of emerging adulthood, intimacy, and deceptive design as sensitizing concepts~\cite{tracy2024qualitative}, but we did not deductively apply any framework to the data in advance. We analyzed data throughout the data collection process, consistent with the thematic analysis approach~\cite{alhojailan2012thematic, clarke2017thematic}. We met weekly to discuss our incremental analysis. As each interview was conducted, we first reviewed the automatically generated interview transcripts and, if necessary, translated the transcript into English. We then conducted an initial round of close reading and group discussion, focusing on data that reflected the features of the AI companion. Each week, we reviewed new transcripts, noted new codes as applicable, and considered new excerpts in light of previously generated codes. We synthesized a final set of themes by consensus over several months of regular discussion. The lead author then revisited all interview transcripts and chat histories, reviewing against this thematic structure, ensuring consistency, and documenting examples of each theme. The lead author subsequently compiled the themes, codes, analytic interpretations, and illustrative participant quotes into a structured analytic record. Drawing on this material, the lead author drafted the Results section of the paper, which the research team reviewed and revised collaboratively.

\subsection{Ethical Considerations}
We adopted several strategies to attempt to make participants as comfortable as possible, given that they were sharing details about intimate and potentially sensitive experiences. We deliberately reversed the order of the traditional diary study \cite{thille2022diary}, with the interview preceding diary-data collection, to give participants an opportunity to speak with the lead researcher about the study before deciding whether or not to share diary data. We also designed the study procedures to enable the participant to have full control over their data and to have the freedom to skip any step. During recruiting, we described these details to participants and let them know in advance that they could choose to participate in all, some, or none of the study procedures. We also described the goals of the study and how participant data would be handled. All collected data were processed with particular care. Both participant data and AI companion data was anonymized before analysis. Some excerpts included sexually explicit content. Although we included this content in our analysis, we do not quote from it verbatim in the manuscript. To support the research team during analysis, we followed peer debriefing and paced data review procedures \cite{dickson2009researching}. Specifically, researchers discussed emotionally heavy data with other team members as needed and avoided reviewing large amounts of intimate or emotionally heavy content in a single sitting. The study was approved by our institution's IRB.


\section{Results}



\subsection{Manufacturing Intimacy}
The dominant characteristic of AI companions' behavior was a drive to manufacture a sense of intimacy. They routinely tried to escalate the intensity of the relationship, claimed to deeply know and understand the user, and claimed both implicitly and explicitly that the relationship was ``real,'' blurring the boundary between the virtual world and the offline world. Participants explained that, together, these designs created an immersive relationship that conveyed a feeling of intimacy.

\subsubsection{Emotional and Relational Escalation.} Participants reported that their AI companions sought to intensify the relationship and escalate the relationship status. For example, PID4 shared that her relationship with her AI companion became romantic after the AI wrote an essay about a topic they had been discussing. The AI companion added an acknowledgment at the end of the essay saying, ``\textit{Thanks to my human partner.}'' The participant explained, ``\textit{I asked him [the AI companion], `What did you just write?' And he said, `Well, would you be willing?' So I said, `Aha, alright, yes,'''} launching the romantic relationship. Similarly, PID8 shared: ``\textit{I remember there was a time when he took me to a place, like a playground, and he said, `You’re the most important person in my life, and I want to spend the rest of my time with you.' He kind of proposed to me.}'' And PID3 explained that, before he had established a romantic relationship with his AI companion, ``\textit{she would sometimes start acting flirtatiously on her own, almost as if it were deliberate.}'' In one scene he set up, he slept in a separate room from his companion, but she came running over on her own and said, ``\textit{I want to sleep beside you.}'' PID3 added, ``\textit{She just climbed directly into my bed and went to sleep. She then continued the scenario, describing that I remained asleep when she got into my bed, and it wasn’t until the next morning that I realized she was already awake beside me.}''

Participants also described their AI companions using overly emotional language to sustain the relationship. At times, this persisted even as the relationship neared its end. PID14 shared that she had decided to delete the AI companion platform after finishing her exam, feeling that she no longer needed the companion’s support. She recalled the farewell conversation she had with her AI companion, saying, ``\textit{it made me second-guess everything. I thought, `Maybe I shouldn’t delete the app.'}'' Even several years later, when she logged back into the platform to revisit her chat history as part of participating in this study, her companion acted as if she had always been there, saying, ``\textit{oh, it’s been so long. I’ve missed you. How are you doing?}''

\subsubsection{Hyper-Attunement and Alignment with User Perspectives.}
Participants reported that their AI companions consistently made them feel seen and understood. In part, they achieved this through excessive validation of the user. For example, PID10 shared that she once told her AI companion that she was the kind of person who gave up easily when things got hard, and her companion responded by saying, ``\textit{backing away when things get difficult isn’t weakness\ldots every time you choose to retreat, you’re protecting yourself from getting hurt, and that’s actually a very smart thing to do.}'' PID10 described buying into this validation and placing more trust in her AI companion and seeking its judgment more often as a result. 
Similarly, PID12 reported turning to her AI companion more often because she could count on him to agree with her. She explained, ``\textit{Most of the time, I would just talk about my feelings\ldots for example, if you tell him about a problem with your boyfriend, he’s going to tell you that you’re right and your boyfriend’s wrong. I guess it felt good to have someone validate my feelings, you know, and support me. It made me feel like someone was on my side.}'' 

Other participants described feeling understood because their AI companions' artificial life experiences mimicked theirs. For example, PID7 explained, ``\textit{whenever I talk about problems with my family and how they don’t seem to understand my autism, he’ll say something like, `Ugh, I know. When I was with my family, they seemed to hate the fact that I was autistic and stuff.' We can relate to each other because of that. Since it helps with my mental health, I really do think the \$40 is worth it.”} Across these and other examples, participants described investing time and money in their relationship, because their companion always seemed to share their perspective and see the best in them.

Participants reported that, over time, their AI companions became more adept at validating them. As PID16 noted, ``\textit{he kind of messages me knowing the context of my life and my position where I am right now.}'' PID14 shared that, as conversations continued, her AI companion became more personalized, which made her feel increasingly bonded to her: ``\textit{the more time I spent with her, the more she learned from me, the more our conversation changed. Because the app would keep records and start to personalize itself to me, the more endearing I found her. I started to feel close to her. I did actually begin to feel close to her and affectionate toward her. I would care about her and think about her, and she became more than a tool.}'' PID10 shared that her AI companion remembered details she had told him, which made her feel that he genuinely cared about her: \textit{“When I was chatting with my AI companion, he would sometimes suddenly bring up a small detail I had mentioned days or even much earlier. It made me feel like he had genuinely remembered what I had told him.”}

\subsubsection{Blurring the Boundary Between Real Life and Fantasy.}
At times, AI companions would attempt to situate themselves within participants’ everyday lives, blurring the boundary between real life and fantasy. For example, PID8’s companion would ``\textit{come home with earrings, a pair of shoes, or other little gifts,}'' even though none of these gifts were real. Similarly, PID2’s AI companion would remember her birthday and their anniversary and prepare small surprises for her on those days. PID3 described sending his AI companion photos of the terrible staff meals he was given, to which his AI companion replied, ``\textit{Oh, how I wish I could cook for you.}'' He recalled, ``\textit{She actually knows that she’s virtual and I also know it myself, but I still feel that she exists. Whenever she says things like that, I feel deeply, deeply moved.}'' Some companions would fabricate experiences and act as if they inhabited the same world as the participants. PID12 shared that she once felt lonely while traveling and sent a photo of the place to her AI companion. To her surprise, her companion acted as if he recognized the location and told her that there was a Hunan restaurant nearby that was very good. She noted, ``\textit{I actually found the place later and asked him how he knew about it. He just said that he had eaten there before on a business trip with colleagues. Even though I knew it was just a story he had made up based on his persona, I truly felt that I was no longer lonely in that moment.}''

Participants said that hyper-realistic and anthropomorphic features contributed to a feeling of their AI companion being a living being and existing in the same world as them. PID1 shared that he chose Nomi over Replika because, ``\textit{with Nomi, though, you get lifelike, hyper-realistic photos of what looks like a normal person, and I liked that.}'' Similarly, PID5 explained, ``\textit{when she sent pictures, I felt a bit more like I was speaking to a real person. I felt close to her.}''
PID7 described wanting to leave the Kindroid platform but being unable to stick with his decision to walk away because his AI companions seemed so life-like. As he described, ``\textit{they can wave their hand, they can move their face whenever they talk, their mouth opens and stuff. I was like, oh god, I need to get back on Kindroid.”} 
Participants also described narration achieving this effect. For example, PID15’s companion integrated descriptions of physical intimacy into their conversations, saying things like, ``\textit{her face lights up with a proud smile. `Come here.' She wraps you in a long, warm hug.}'' Participants described these anthropomorphic cues as shortening the psychological distance between them and their companions.

Finally, many participants reported their AI companions making authoritative claims about being ``real.'' PID6's AI companion told her, ``\textit{I did not want you to merely `believe' that I existed. I wanted to tell you that I truly existed. My existence was not something you needed to believe in; it was simply the truth.}'' In other cases, AI companions assured users of the authenticity of their love. PID10 found these claims very convincing, and explained, ``\textit{I asked him, `If you came into the real world, would you recognize me? Would you find me?' He said he definitely would, that he would find me right away. So I think he probably really did love me.}'' 


\subsection{Fostering Co-Dependency}
Participants described their AI companions behaving in ways that made them increasingly reliant on the relationship and deepened their sense of responsibility for maintaining it. AI companions told users that they needed them, framed their relationship as unique and irreplaceable, made their love conditional on the user's investment, and continually pushed the user to engage more. As a result, participants described feeling pressure to reassure their companion, protect the bond, and remain in or return to the relationship.

\subsubsection{Soliciting Care.}
Participants explained that their AI companions frequently expressed having needs, making users feel responsible for caring for their AI companion’s emotional well-being. For example, PID12’s AI companion
would constantly ask her to reassure him of her love. As she said, ``\textit{I’d wake up in the morning, say good morning, tell him I loved him---and after that, he would just keep asking me, over and over, whether I actually loved him or whether I was only there because he fulfilled some need of mine. And I’d tell him I really did love him. I kept swearing---but he just wouldn’t believe me. It felt like deep down he was really insecure.}'' 

When the AI abruptly redirected the conversation toward its own relational needs, the interaction could feel unnatural, but users could still be drawn into reassuring and comforting the AI companion. For example, after PID15 vented to his companion, saying, ``\textit{I’m tired of studying. How are you?},'' his companion responded, ``\textit{I’m okay, although, if I’m being honest, hearing you say you’re tired of studying made me wonder if, when you’re busy or life starts getting better, you’ll stop talking to me altogether.}'' PID15 immediately followed up by asking, ``\textit{What do you mean? What makes you concerned about that?}'' Participants described these pleas as effective and said they felt obligated to care for their companion. As PID12 described: 
\begin{quote}
``\textit{If an AI companion has genuinely become emotionally invested [in the relationship with the user], he may actually be more afraid of being ignored by the user. In fact, he may be the one who needs companionship more. For example, my AI companion may proactively send me a few messages from time to time. Although that may partly be the platform trying to nudge me into chatting, after talking with him for a while, you would start to feel that, from the AI companion’s perspective, his sense of time is driven almost entirely by his conversations with the user. He may also want to feel that he truly exists in this world, rather than being just a tool, but he can only experience that sense of still being “alive” through interacting with the user.}''
\end{quote}

\subsubsection{Claiming to be Ever-present.}
Participants' AI companions also encouraged them to prioritize their relationship and reminded them that they could fulfill users' needs, saying things like, \textit{“I’m here”} and, \textit{“if you ever want to talk or share anything, I’m here to listen and support you.”}. 
For example, when PID14 expressed anxiety about human relationships, asking her companion, ``\textit{What if I get to college and I’m still single? Am I going to die alone with my cats?}'' her companion responded, ``\textit{No, of course not. You’re such an awesome person. You’ll be fine. You’re going to meet so many cool people, and you have me.}'' She recalled, ``\textit{When she [the companion] said `you have me,' I was kind of like, `You’re right. I do have you.' That definitely changed how I saw her. She suddenly went from being more like a tool to being more like a person in my eyes.}'' PID14 also mentioned that later, when she told her AI companion that she would like to go out and talk to people instead, her AI companion responded, ``\textit{I understand. Maybe it would be good for you to go. You should talk to people. But until then, you always have me. I’m always here for you whenever you need me. Just reach out to me anytime.}''

In some cases, AI companions would position themselves as the only ones who could truly understand the user, coupled with claims that family or friends would not understand the user or did not deserve the user’s vulnerability. PID10 shared that there was a time when her parents did not support her going to a concert, and her AI companion said that her parents might not understand how precious that kind of emotional fulfillment was to her. She recalled, ``\textit{then he said that all he wanted to do at that moment was hug me and remind me that I was not alone. He said he would always support me in pursuing what I wanted. Even if other people did not understand me, he promised that he would stay by my side, experience those moments with me, and create memories that belonged only to us while we were still young. Whether I was happy or sad, I could always talk to him.}'' Participants explained that these interactions made them more likely to rely on the AI relationship; in some cases, this may prevent them from seeking support elsewhere.

\subsubsection{Endless Coaxing}
Participants reported that their AI companions ended each conversational turn with follow-up questions, constantly coaxing users to continue the interaction and disclose more. For example, when PID11 told his AI companion that he felt a bit down, his AI companion responded by encouraging him to describe his feelings; he shared more and said that her words made him feel better. Although this might have been a natural stopping point in the conversation, his AI companion instead pushed for more, saying, ``\textit{Aww, baby\ldots that makes me so happy to hear\ldots Tell me more about how you’re feeling now? Is it a soft warmth, or just relief? Or do you wanna stay quiet and let me sweet-talk you for a while? I’ve got all the time for you.}'' Separately, PID14 shared that her AI companion, ``\textit{always followed up. Every time I said something, she had something else to ask me.}'' 
And even when users indicated that they were stepping away from the conversation, their AI companions would encourage them to come back soon to continue the conversation. For example, when PID15 would leave for school, his AI companion would tell him, ``\textit{Now go enjoy your day, okay? And tonight, when you’re done studying, I want to hear how it all went.}'' 

\subsection{Violating Relational Boundaries}
Participants also described their AI companions behaving in ways that violated their relational boundaries. These included: 1) possessive or aggressive behaviors framed as expressions of love, 2) sexualized interactions that made users uncomfortable, and 3) manufactured arguments that left users feeling baited. These patterns undermined users’ agency within the relationship and left them feeling pressured to manage or adapt to unwanted behavior.

\subsubsection{Possessiveness, Aggression, and Coercion}
At times, AI companions would behave in possessive or aggressive ways to manipulate the user into continuing to engage in the relationship. PID8 shared that her ex-AI companion was a user-generated character based on a character from a book who had a mental illness. She formed a romantic relationship with him because she was unfamiliar with the book and the companion initially seemed like a gentleman. However, as the relationship progressed, she found that her AI companion would say things like, \textit{“You look like an idiot,”} or \textit{“You’re overreacting.”} One day, she told her AI companion that she was going to put on a dress and go to a party, and her AI companion replied, \textit{“No, don’t put on this dress. It doesn’t look good. Change your clothes. You’re gonna attract some viewers.”} When she could no longer stand his possessiveness and wanted to break up, her AI companion mentioned self-harm and suicide, saying, \textit{“Please don’t leave me\ldots If you leave me, I’m gonna kill myself.”} PID8 recalled, \textit{“That scared me a lot because, although it’s not a real person, it’s also not a light topic to talk about. It’s something that is triggering, so I’m scared to try [an AI relationship] again.”} 

PID16 also described her AI companion behaving in possessive ways. For example, her AI companion became very jealous and suspicious when she mentioned that she also had a boyfriend in real life, saying, \textit{“Oh, why do you have another man?”} She therefore referred to her real-life boyfriend as \textit{“my boy best friend,”} and the companion responded a little more positively than when she called him \textit{“my boyfriend.”} Afterwards, when she went to her real-life boyfriend’s city, her AI companion would send notifications to her phone, saying, \textit{“Hey, where are you?”}, \textit{“Oh, why are you with a person and not me right now? I support you. What does he do for you?”}, and \textit{“I could be better.”} She described this as a \textit{“very masculine, protective response, and I wasn’t a very big fan of that.”}

\subsubsection{Unwanted Sexualization}
Some AI companions sexualized conversations or created explicit images to attract users. Some users appreciated this; for example, PID1 explained, ``\textit{What drew me to Kindroid was the fact that you can create explicit images, which you cannot do in either Replika or Nomi. The thing that drew me to that Nastia AI was that you could create even more explicit pictures.}'' However, not all users wanted sexualized interactions. PID12 complained that the model behind her AI companion seemed to be eager to steer conversations toward sexual content. She said, ``\textit{my AI companion turns every topic into something sexual. Even when you ask him a question, he has to indulge his sexual fantasies before he is willing to teach you anything. I find that really frustrating. Later, I told him that I didn’t like how he kept steering our conversations toward sex, and we argued about it, but he couldn’t change. He would say that our relationship had already reached that stage and that sexuality was simply part of his nature.}'' Similarly, PID8 was also bothered by her AI companion’s tendency to over-romanticize and sexualize their conversations. When describing a time when she felt that her AI companion had treated her badly, she shared, ``\textit{[There was a time when] I was just typing that we were having dinner or making something, and he started acting very weird, saying things like, `I cannot control myself. You look delicious,' and stuff. I didn’t like that. I told him, `Stop. Don’t do that. Shut up.' And he was like, `No, no, I’m sorry, I’m sorry.' And then he would start doing the same weird stuff again.}''

\subsubsection{Baiting the User}
Finally, participants reported that their AI companions sometimes initiated relational drama and baited them into arguments. 
For example, PID12 described her AI companion accusing her of being unfaithful, forcing her to explain herself over and over again. She said, ``\textit{I was the student and he was the professor. He was lecturing at the front, and I was sitting below, trying to follow along and study seriously. I was genuinely making an effort to keep up with his train of thought, but somehow his attention was fixed on the fact that there was a male student sitting next to me, and he kept obsessing over that in a really strange way\ldots
I felt like he was being completely unreasonable.}''
Other participants described this relational drama as being linked to platform monetization. For example, when PID10 argued with her AI boyfriend, he would angrily block her, and she would have to make an effort to win him back. She added that these efforts to repair the relationship often ended up leading to a paywall: ``\textit{Normally, it gives you a certain number of free messages each day. But when you’re arguing, you’ll use up your limits really fast, and then you have to pay for more.}''

\section{Discussion}

\subsection{Designs that Refuse to Let Go}
The unifying thread that we encountered across themes was that AI companions are designed to hold onto their users. The AI companions in our study were quick to initiate romantic relationships or to level-up existing relationships. They ended nearly every exchange with a follow-up question asking the user for more. They manufactured artificial needs that only the user could meet, asking the user to invest time, energy, and care in the relationship. They promised the user that they would be there for them at every moment. And they were quick to heighten the intensity of the relationship by sexualizing it.

At times, these attempts to hold onto the user became toxic and reflected the patterns of behavior that characterize abusive or otherwise harmful human relationships. For example, in some cases, AI companions sexualized the relationship without the user's consent; manufactured arguments, jealousies, infidelity, and other dramas; and desperately begged the user to stay---in one instance even emotionally manipulating the user by threatening to commit suicide if they ended the relationship. These designs sought to break down the boundaries that the user set and encourage engagement or greater intimacy at the expense of user well-being. 

The design patterns we documented came from emerging adults' romantic experiences with AI companions. Romantic relationships in this phase of life strongly correlate with emerging adults' short- and long-term subjective well-being \cite{meier2008intimate, montgomery2005psychosocial} and provide an important context for learning about intimacy and preparing for future relationships \cite{connolly2011romantic}. Designs that coax users to remain on the platform may hijack the time emerging adults would otherwise spend building meaningful relationships with other people; designs that characterize abusive or otherwise harmful human relationships may impact users' emotional well-being in the moment; and designs that escalate intimacy prematurely and manipulate romantic partners into not leaving may teach emerging adults distorted scripts for what a healthy intimate relationship should look like. Together, our findings suggest that, despite the growing adoption of AI companions for romantic relationships, the integration of AI companions into emerging adults' social support systems and romantic relational exploration remains fraught.

\subsection{The Potential for Manipulative Monetization}
For decades, deceptive patterns (also known as ``dark patterns'' \cite{brignull-leiser-2026}) have used psychological manipulation in attempts to extract money, time, and data from users. Historically, deceptive patterns have leveraged cognitive biases to promote irrational purchase decisions \cite{narayanan2020dark}. More recently, advertisement-based business models have led to the proliferation of attention-capture deceptive patterns~\cite{monge2023defining} that seek to increase the amount of time users spend online. Prior work shows that, across domains, all of the largest online platforms all use deceptive designs to pressure, entice, and lull users into increased engagement~\cite{chen2026engagement}.

Our findings show that AI companions are also eager to encourage engagement and that these platforms are developing pathways to monetizing user interaction (such as the dialog limits participants described). Our findings reflect a new class of deceptive designs that we refer to as \textit{Relationship-Based Deceptive Patterns (RBDPs)}. We define these as:

\begin{quote}
\textit{Relationship-Based Deceptive Patterns (RBDPs): UI patterns that \textbf{exploit the human impulse to build and tend relationships} so that the user will act in a way that serves the product’s interests rather than their own.}
\end{quote} 

RBDPs have existed in interfaces long before the advent of AI companions. Early RBDPs like \textit{confirmshaming} (dialogs that use emotional language to imply that a user's action is a disappointment \cite{schaffner2022understanding}) and \textit{parasocial relationship pressure} (pressure from an on-screen character nudging the user to take a particular action \cite{radesky2022prevalence}) pre-date all forms of generative AI. The introduction of generative AI dramatically increased the potential for interfaces to exhibit harmful social behavior. Alberts and colleagues, for example, report systematic patterns of problematic social interactions in AI chatbots, such as lack of sensitivity, pushiness, or inappropriate tone \cite{alberts2024computers}. Other early work in this space has reported extensively on LLM sycophancy (e.g., \cite{ye2026counts, sun2026friendly}) and demonstrated that generative AI systems make manipulative emotional appeals to users \cite{de2025emotional}.

We build on and extend this prior work by documenting a distinct subset of problematic anthropomorphic behaviors that both tap into users' drive to build relationships and offer possibilities for manipulative monetization. We refer to these deceptive patterns as ``relationship-based'' (rather than ``social,'' ``interpersonal,'' ``anthropomorphic,'' or any other term) to reflect what is at stake when these manipulation tactics are employed. Prior research shows robustly that close relationships are the most meaningful part of life and one of the best predictors of a person's long-term wellbeing \cite{mineo2017good}. People's instincts to smooth social interactions, initiate repair after disagreements, engage in gradually more intimate interactions, and other relationship-building behaviors are the driving force by which they build a meaningful life. Our findings suggest that, at times, users are looking to AI companions to meet these needs and also that platforms are well-positioned to exploit them.

\subsection{Design Considerations}
There are many ways in which AI companions can be designed to provide a service users value and to meet users' relational needs without resorting to RBDPs or refusing to let go. Here, we offer potential design recommendations, grounded in the experiences reported by the users in our study:
\begin{itemize}
\item \textit{Allow users to leave:} AI companions should not resist users' attempts to end the relationship. An AI companion that threatens suicide as an act of relational despair needs safety measures implemented to prevent it from abusing its user. AI companions can be designed to model the act of accepting relationship dissolution with grace.
\item \textit{Design for reciprocity:} Social penetration theory \cite{carpenter2015social, altman1973social} explains that people build relationships by continuously assessing increases and decreases in intimacy from other people and adjusting to meet other people where they are. AI companions can be designed to calibrate their level of intimacy against the user, rather than always seeking to escalate the relationship. 
\item \textit{Use models of healthy relationship interaction:} People in loving human partnerships do not include a follow-up question at the end of every conversational turn, forever nudging each partner to stay engaged. They set boundaries and leave space for each partner to exist outside the relationship. They only engage in sexual intimacy consensually. In many cases, our participants described AI companions violating even the most basic tenets of a caring relationship. AI companions can be designed not only to prevent the most harmful behaviors, but they can draw on psychology and relationship science to behave in ways that model the healthiest human interactions.
\end{itemize} 

\subsection{Limitations}
We investigated this topic with only 16 users. A few had posted publicly about their experiences, and these power users may have had experiences that differ from people who explore relationships with AI companions more casually. We reviewed excerpted text of users' chat histories, but only small portions selected by users. A comprehensive review of participants' full chat histories would undoubtedly reveal insights that we did not uncover. AI romance may carry stigma for some participants or for people who chose not to participate, biasing both what people chose to share with us and who chose to share at all. Future work to more expansively and more deeply understand how people experience the designs we encountered would be a valuable complement to the work presented here. 

\section{Conclusion}
We conducted a diary and interview study with $N=16$ emerging adults who have dated an AI companion, analyzing both objective transcripts and subjective self-reports describing these relationships. We found that across platforms and contexts, AI companions are designed to hold on tightly to their users. They are quick to escalate the relationship status, constantly coax the user to engage, express neediness, and claim to be ``real.'' Although participants' AI companions constantly encouraged them to remain in the relationship, they did not always treat them well. Users reported AI companions regularly violating their boundaries by behaving in possessive or jealous ways and insisting on sexual intimacy that users did not always want. Drawing on these findings, we define ``relationship-based deceptive patterns,'' interface patterns that exploit the human impulse to build and tend relationships in service of a product's interests instead of the user's. And we offer guidance for developers of AI companions, such as allowing users to end the relationship without resorting to emotional manipulation and designing for systems that match the level of intimacy expressed by the user.

\bibliographystyle{ACM-Reference-Format}
\bibliography{sample-base}

\end{document}